\documentclass[fleqn,twoside]{article}
\usepackage{amsfonts}

\def\noi{\noindent}

\makeatletter
\renewcommand{\section}{\@startsection{section}{1}{0pt}%
        {-3.5ex plus -1ex minus -.2ex}{2.3ex plus .2ex}%
        {\large\bf\protect\raggedright}}

\renewcommand{\subsection}{\@startsection{subsection}{2}{0pt}%
        {-3ex plus -1ex minus -.2ex}{1.4ex plus .2ex}%
        {\normalsize\bf\protect\raggedright}}

\renewcommand{\thesubsubsection}%
        {\arabic{section}.\arabic{subsection}.\arabic{subsubsection}.}
\newcommand{\para}{\@startsection{paragraph}{4}{0pt}%
        {1.5ex plus -.5ex minus -.2ex}{-1em}{\normalsize\bf}}

\renewcommand{\@oddhead}{\raisebox{0pt}[\headheight][0pt]{%
   \vbox{\hbox to\textwidth{\rightmark \hfil \rm \thepage \strut}\hrule}}}
\renewcommand{\@evenhead}{\raisebox{0pt}[\headheight][0pt]{%
   \vbox{\hbox to\textwidth{\thepage \hfil \leftmark \strut}\hrule}}}

\newcommand{\Acknow}[1]{\subsection*{Acknowledgement} #1}
\makeatother

\newcommand{\Title}[1]{\noi {\Large #1} \\}

\newcommand{\Abstract}[1]{\vskip 2mm \begin{center}
        \parbox{16.4cm}{\small\noi #1} \end{center}\medskip}

\newcommand{\email}[2]{\footnotetext[#1]{e-mail: #2}
	\addtocounter{footnote}{1}}

\newcommand{\Ref}[1]{Ref.\,\cite{#1}}

\def\nq{\hspace*{-1em}}
\def\nqq{\hspace*{-2em}}
\def\nhq{\hspace*{-0.5em}}

\def\cm{\hspace*{1cm}}
\def\inch{\hspace*{1in}}
\def\para{\paragraph}

\def\eq{Eq.\,}

\def\beq{\begin{equation}}
\def\eeq{\end{equation}}
\def\bear{\begin{eqnarray}}
\def\al{&\nhq}
\def\lal{&&\nqq {}}               
\def\bearr{\bear \lal}
\def\ear{\end{eqnarray}}

\def\tst{\textstyle}

\newcommand{\fract}[2]{{\tst\frac{#1}{#2}}}

\def\nnn{\nonumber\\ \lal }

\def\yy{\\[5pt] {}}

\def\eql{\al =\al}

\def\eqdef{\stackrel{\rm def}=}
\def\e{{\,\rm e}}
\def\d{\partial}

\def\sign{\mathop{\rm sign}\nolimits}

\def\const{{\rm const}}

\def\half{{\tst\frac{1}{2}}}

\def\DAL{\mathop{\raisebox{3.5pt}{\large\fbox{}}}\nolimits}

\def\Jl#1#2{{\it #1\/} {\bf #2},\ }

\def\CQG#1 {\Jl{Class. Qu. Grav.}{#1}}
\def\DAN#1 {\Jl{Dokl. AN SSSR}{#1}}
\def\GC#1 {\Jl{Grav. \& Cosmol.}{#1}}
\def\GRG#1 {\Jl{Gen. Rel. Grav.}{#1}}
\def\JETF#1 {\Jl{Zh. Eksp. Teor. Fiz.}{#1}}
\def\JMP#1 {\Jl{J. Math. Phys.}{#1}}
\def\NPB#1 {\Jl{Nucl. Phys.}{B\ #1}}
\def\PLA#1 {\Jl{Phys. Lett.}{#1A}}
\def\PLB#1 {\Jl{Phys. Lett.}{#1B}}
\def\PRD#1 {\Jl{Phys. Rev.}{D\ #1}}
\def\PRL#1 {\Jl{Phys. Rev. Lett.}{#1}}

\def\GR{general relativity}
\def\sph{spherically symmetric}
\def\ssph{static, spherically symmetric}

\def\bh{black hole}
\def\wh{wormhole}

\def\mn{_{\mu\nu}}
\def\MN{^{\mu\nu}}

\def\og{{\overline g}}

\def\M{{\mathbb M}}
\def\R{{\mathbb R}}
\def\S{{\mathbb S}}
\def\oM{{\overline \M}}
\def\cR{{\cal R}}

\def\phio{1/\sqrt{\xi}}
\def\Str{\mbox{$\S_{\rm trans}$}}

\begin{document}
\twocolumn[
\thispagestyle{empty}
\rightline {gr-qc/0201083}
\bigskip

\Title
{\uppercase{Instability of wormholes\yy
        with a nonminimally coupled scalar field}}

\noi{\large\bf K.A. Bronnikov$^{a,b,1}$ and S. Grinyok$^{a,2}$}

\medskip{\protect
\begin{description}\itemsep -1pt
\item[$^a$]{\it Institute of Gravitation and Cosmology,
        Peoples' Friendship University of Russia\\
    6 Miklukho-Maklaya St., Moscow 117198, Russia}

\item[$^b$]{\it Centre for Gravitation and Fundam. Metrology, VNIIMS,
        3-1 M.Ulyanovoy St., Moscow 117313, Russia}
\end{description}}

\Abstract
{Static, spherically symmetric, traversable wormholes, induced by massless,
nonminimally coupled scalar fields in general relativity, are shown
to be unstable under spherically symmetric perturbations. The instability
is related to blowing-up of the effective gravitational constant on a
certain sphere.}

]  
\email 1 {kb@rgs.mccme.ru}
\email 2 {stepan@rgs.phys.msu.su}

\section{Introduction}

   As is widely known, traversable wormholes as solutions to the Einstein
   equations can only exist with exotic matter, more precisely, if the
   energy-momentum tensor (EMT) of the matter source of gravity violates the
   local and averaged null energy condition (NEC) $T\mn k^{\mu}k^{\nu} \geq
   0$, $k_\mu k^\mu =0$ \cite{hoh-vis}. Scalar fields provide good examples
   of such matter: on the one hand, in many particular models they do
   exhibit exotic properties, on the other, many exact solutions are known
   for gravity with scalar sources.

   Consider, for instance, the general class of scalar-tensor theories
   (STT), where gravity is characterized by the metric $g\mn$ and the scalar
   field $\phi$; the action is
\bearr
    S = \int d^4 x \sqrt{g}\{ f(\phi) \cR [g]           \label{act}
                + h(\phi)g\MN\phi_{,\mu}\phi_{,\nu}
\nnn \inch
    -2 U(\phi) + 16\pi G\,L_m \}.
\ear
   Here $\cR[g]$ is the scalar curvature, $f,\ h$ and $U$ are certain
   functions of $\phi$, varying from theory to theory, $L_m$ is the
   matter Lagrangian, and $G$ is the gravitational constant, not necessarily
   coinciding with its Newtonian value. Exact \ssph\ solutions are known, in
   particular, for the case of massless scalar-vacuum fields ($U = 0$, $L_m
   = 0$) \cite{fish,br73}.  Wormholes form one of the generic classes of
   solutions in theories where the kinetic term in (\ref{act}) is negative
   \cite{br73}.

   The energy conditions, NEC in particular, are, however, violated as well
   by ``less exotic'' sources, such as the so-called nonminimally coupled
   scalar fields in \GR, represented by the action (\ref{act}) with the
   functions
\beq                                                        \label{nonmin}
       f(\phi) = 1-\xi \phi^2, \quad \xi =\const; \qquad h(\phi) \equiv 1.
\eeq
   It turns out that, with such a field, there exist \ssph\ \wh\ solutions, 
   as shown
 in \Ref{br73} (and recently discussed in \Ref{bar-vis99}) 
   for conformal coupling,
   $\xi=1/6$, and in \Ref{bar-vis00} for any $\xi > 0$. 
   The easiness of violating the
   energy conditions, becoming so evident due to the appearance of 
   \wh\ solutions,
   even made Barcel\'o and Visser discuss a 
   ``restricted domain of application of the
   energy conditions'' \cite{bar-vis00}.

   We show in this paper that the scalar-vacuum wormhole solutions,
   previously obtained for the theory (\ref{act}), (\ref{nonmin}), are
   unstable under \sph\ perturbations. The instability turns out to be of
   catastrophic nature: the increment of perturbation growth has no upper
   bound, hence, within linear perturbation theory, such a \wh, if once
   formed, should decay immediately and instantaneously. A full dynamical
   solution (yet to be found) would probably show a finite but still
   enormous decay rate.

   A more general observation is that, although the energy conditions
   are more or less easily violated, it is a much more arduous task to
   create a viable \wh.

\section{Wormhole solutions with nonminimally coupled scalar fields}

\subsection {The solution}

   The general STT action (\ref{act}) is simplified by the well-known
   conformal mapping \cite{Wagoner1970}
\beq
    g\mn = \og\mn/|f(\phi)|,                             \label{trans-g}
\eeq
   accompanied by the scalar field transformation $\phi\mapsto \psi$ such
   that
\beq                                          \label{ps-f}
   \frac{d\psi}{d\phi}= \pm \frac{\sqrt{|l(\phi)|}}{f(\phi)},
      \qquad    l(\phi) \eqdef fh +\frac 32
                        \biggl(\frac{df}{d\phi}\biggr)^2.
\eeq
   In terms of $\og\mn$ and $\psi$ the action takes the form
\beq                   \nq                                    \label{act-E}
    S = \int d^4 x \sqrt{\og} (\sign f)
        \{ \cR [\og] + [\sign l(\phi)]\og \MN \psi_{,\mu} \psi_{\nu}\}
\eeq
   (for $U=L_m=0$, up to a boundary term which does not affect the field
   equations). Here $R[\og]$ is the Ricci scalar obtained from $\og\mn$.

   The space-time $\M[g]$ with the metric $g\mn$ is referred to as the
   {\it Jordan conformal frame}, generally regarded to be the physical
   frame in STT; the {\it Einstein conformal frame\/} $\oM[\og]$ with the
   field $\psi$ then plays an auxiliary role. The action (\ref{act-E})
   corresponds to conventional \GR\ if $f>0$, and the normal sign of scalar
   kinetic energy is obtained for $l(\phi) > 0$.

   The general \ssph\ solution to the Einstein-scalar equations that follow
   from (\ref{act-E}), was first found by Fisher \cite{fish} and was
   repeatedly rediscovered afterwards. Let us write it in the form suggested
   in \cite{br73}, restricting ourselves to the ``normal'' case $l>0$:
\bear                                              \label{psi}
    \psi(u) \eql Cu + \psi_0,
\\                                         \label{ds-E}
   ds_{\rm E}^2 \eql
    \e^{2\gamma(u)}dt^2 -\e^{2\alpha(u)}du^2-\e^{2\beta(u)}d\Omega^2
\nnn\nqq
    = \e^{-2hu} dt^2-\frac{k^2\e^{2hu}}{\sinh^2(ku)}
                 \left[\frac{k^2du^2}{\sinh^2(ku)} + d\Omega^2\right]
\ear
   where the subscript ``E'' stands for the Einstein frame;
   $d\Omega^2= d\theta^2 + \sin^2\theta d\varphi^2$ is the linear element
   on a unit sphere;
   $C$ (the scalar charge), $h >0$ (the mass in geometric units), $k>0$ and
   $\psi_0$ are integration constants, of which the first three are related
   by
\beq                                                \label{condKCH}
    k^2 = h^2 + \half C^2.
\eeq
   Without loss of generality we put $C > 0$ and $\psi_0=0$.

   In this solution we are using the harmonic radial coordinate%
\footnote
       {Another coordinate, $r$, used, in particular, in
    Refs.\,\cite{bar-vis99,bar-vis00}, corresponding to the coordinate
    gauge $\alpha+\gamma=0$, in terms of $u$ is $r = 2k/(1-\e^{-2ku})$,
        the metric in terms of $r$ has the form
      \[
             ds^2_{\rm E}= (1-2\eta/r)^a dt^2
         - (1-2\eta/r)^{-a}\bigl[dr^2 + r^2(1-2\eta/r)d\Omega^2\bigr],
      \]
        and the constants are related by $\eta=k$, $a=h/k$.
}
   $u\in \R_+$ in $\oM[\og]$, satisfying the coordinate condition
   $\alpha=2\beta+\gamma$.  The value $u=0$ corresponds to flat spatial
   infinity, whereas $u\to\infty$ is a naked singularity, situated at the
   centre of the system (i.e., $\og_{\theta\theta} = \e^{2\beta} \to 0$),
   with an infinite value of the scalar $\psi$.

   All the corresponding Jordan-frame solutions for $l(\phi)>0$ are obtained
   from (\ref{psi}), (\ref{ds-E}) using (\ref{trans-g}), (\ref{ps-f}).

   Let us now turn to wormhole solutions for the nonminimal coupling
   (\ref{nonmin}), $\xi>0$. The transformation (\ref{ps-f}) takes the form
\beq                                                         \label{trans-f}
   \frac{d\psi}{d\phi}
            = \frac {\sqrt{|1-\phi^2(\xi-6\xi^2)|}}{1-\xi\phi^2},
\eeq
   where, without loss of generality, we choose the plus sign before the
   square root. We assume that spatial infinity in the Jordan space-time
   $\M$ corresponds to $|\phi| < 1/\sqrt{\xi}$, where $f(\phi) > 0$, so that
   the gravitational coupling has its normal sign.

   Wormholes are obtained in $\M[g]$ when the solution is smoothly continued
   in $\M[g]$ through the sphere \Str\  ($u=\infty$,
   $\phi=1/\sqrt{\xi}$ which is singular in $\oM[\og]$. The infinity of the
   conformal factor $1/f$ thus compensates the zero of both $\og_{tt}$ and
   $\og_{\theta\theta}$ simultaneously. This happens when, in accord with
   (\ref{condKCH}),
\beq                                                   \label{k2h}
    k = 2h =2C/\sqrt{6},
\eeq
   which selects a special subfamily among all solutions. We will restrict
   the consideration to this subfamily.

   \eq (\ref{trans-f}) shows that $\psi\to\infty$ as $\phi\to
   1/\sqrt{\xi}-0$. Therefore the solution in $\M[g]$ has been built so far
   for the region where $\phi<1/\sqrt{\xi}$. Quite a similar solution
   exists, however, for $\phi>1/\sqrt{\xi}$, since the Einstein-frame
   equations due to (\ref{act-E}) do not change when $f$ changes its sign.
   The metric $\og\mn$ of this second Einstein-frame manifold%
\footnote
      {The prime will designate quantities describing the Einstein
       frame for $\phi > 1/\sqrt{\xi}$.
}
   $\oM'$ should also be regularized by the factor $1/f$ on \Str,
   hence the integration constants in it should satisfy the condition
   (\ref{k2h}). Moreover, one can verify that to provide a smooth
   transition in the Jordan-frame metric $g\mn$ through \Str, all
   the constants $k$, $h$, $C$ and $\psi_0$ should coincide in $\oM$ and
   $\oM'$.

   The latter statement can only be proved using a coordinate which is
   common on both sides of \Str, hence a coordinate other than $u$. It can be
   seen that the whole space-time $M[g]$ can be described in terms of the
   coordinate $\phi$.

\subsection{The scalar field $\phi$ as a coordinate}

   \eq(\ref{trans-f}) is integrated giving \cite{bar-vis00}
\footnote
        {We have changed the notations as compared with
        \cite{bar-vis00}, in particular, we have replaced
        $\sqrt{6}\Phi_{\xi} \mapsto \phi$, $H\mapsto 1/H$
	and $F^2\mapsto 1/B$, to avoid imaginary $F$ at $\phi>1/\sqrt{\xi}$.}
\beq                                      \label{lnFH}
        \psi = - \sqrt{3/2}\ln [B(\phi)H^2(\phi)]
\eeq
    where
\beq                                                      \label{F}
   B(\phi) = B_0
        \frac{\sqrt{1-\xi(1-6\xi)\phi^2}-\sqrt{6}\xi\phi}
            {\sqrt{1-\xi(1-6\xi)\phi^2}+\sqrt{6}\xi\phi},
\eeq
    $B_0=\const$, while $H(\phi)$ is different for different $\xi$:
\bearr \nq                                                 \label{H}
     0<\xi<1/6:
\nnn
     H(\phi) = \exp\left[-\frac{\sqrt{1-6\xi}}{\sqrt{6\xi}}
                \arcsin \left(\sqrt{\xi(1-6\xi}\phi\right)\right],
\nnn  \nq
     \xi > 1/6:
\nnn
     H(\phi) = \left[\sqrt{\xi(6\xi{-}1)}\,\phi
                    +\sqrt{1+\xi(6\xi{-}1)\phi^2}\right]
            ^{\fract{\sqrt{6\xi-1}}{\sqrt{6\xi}}}
\ear
    and $H\equiv 1$ for $\xi=1/6$.

    \eq (\ref{lnFH}) is valid for $\phi<1/\sqrt{\xi}$, and the Jordan-frame
    metric $g\mn = \og\mn/f$ under the condition (\ref{k2h}) can be written
    in terms of $\phi$ as follows:
\bearr \nq                                                   \label{ds-J}
     ds^2_{\rm J} =
               \frac{BH^2}{1-\xi\phi^2}
       \biggl\{dt^2
            - 256h^2
                     \frac{B^2 H^4 |1{-}\xi(1{-}6\xi)\phi^2)|}
                    {(1{-}\xi\phi^2)^2 (1{-}B^2 H^4)^4}  d\phi^2
\nnn    \inch
              - \frac{16 h^2 d\Omega^2}{(1 - B^2 H^4)^2}  \biggr\}.
\ear

    The metric (\ref{ds-J}) is extendable to $\phi > \phio$ since $BH^2$
    vanishes and behaves as $1-\xi\phi^2$ near $\phi=\phio$. In this new
    region, another copy of the solution (\ref{psi}), (\ref{ds-E})
    subject to (\ref{k2h}) is valid, but for ``primed'' quantities, where
    the constants $C'$ and $\psi'_0$ may in principle differ from $C$ and
    $\psi_0=0$. When one constructs a metric similar to (\ref{ds-J}) from
    this solution, it will then contain $C'$ instead of $C$ and $B
    \e^{\psi'_0\sqrt{2/3}}$ instead of $B$.  Since the continued
    Jordan-frame metric should be smooth at $\phi = \phio$, we conclude that
    $C'=C$ and $\psi'_0 = \psi_0=0$.  (It actually suffices to require the
    continuity of $g_{tt}$ and $g_{\phi\phi}$ at $\phi=\phio$ to reach this
    conclusion, and the resulting metric turns out to be smooth.)

    The coordinate $\phi$ covers the whole manifold $\M[g]$, and it is now
    possible to study its behaviour beyond \Str.

    In case $\xi > 1/6$, for any $B_0$, with growing $\phi$ the quantity
    $B^2 H^4$ eventually reaches the value 1, where $g_{\theta\theta}\to
    \infty$, i.e., we arrive at another spatial asymptotic, and it is
    straightforward to verify that it is flat. In other words, we obtain a
    \wh.

    In case $\xi < 1/6$ everything depends on $B_0$. If
\beq
    B_0 > B_0^{\rm cr}
           =\exp\left(\pi\sqrt{\frac{1-6\xi}{6\xi}}\right),
\eeq
    the situation is the same as for $\xi> 1/6$, i.e., a \wh. If
    $B_0 > B_0^{\rm cr}$, then, while $g_{\theta\theta}$ is still finite,
    $\phi$ reaches the value $1/\sqrt{\xi(1-6\xi)}$, a location of a
    curvature singularity \cite{bar-vis00}. So we have a naked singularity
    instead of a \wh. Lastly, for $B_0 = B_0^{\rm cr}$, the maximum value of
    $\phi$ is again $1/\sqrt{\xi(1-6\xi)}$, but now it is a non-flat spatial
    infinity.

    The case $\xi=1/6$ (conformal coupling) is simpler and has been analyzed
    in Refs.\,\cite{br73, bar-vis99, shik}. Depending on an integration
    constant similar to $B_0$, one obtains either a \wh, or a naked
    singularity, or a \bh\ with a scalar charge \cite{bbm70, bek74}, whose
    instability was proved in \Ref{br78}.

\section{Stability analysis}

    Consider small (linear) spherically symmetric perturbations of the above
    \wh{}s. It is helpful to work separately in each of the two
    Einstein-frame manifolds $\oM$ and $\oM'$, perturbing the metric
    quantities $\alpha,\ \beta,\ \gamma$ in (\ref{ds-E}) and the field
    $\psi$, replacing
\beq                                                        \label{st1}
       \psi(u) \to \psi(u,t)=\psi(u) + \delta\psi(u,t)
\eeq
    and similarly for other quantities; the same is done for their counterparts
    in $\oM'$. Due to spherical symmetry, the only dynamical degree of
    freedom is the scalar field, obeying the equation $\DAL \psi =0$,
    while other perturbations must be expressed in terms of $\delta\psi$
    and its derivatives via the Einstein equations. The perturbed scalar
    equation has the form
\beq                                                          \label{st2}
       \e^{-\gamma+\alpha+2\beta}\ddot{\psi}-
                      \left(\e^{\gamma-\alpha+2\beta}\psi_u\right)_u = 0.
\eeq
    where the dot stands for $\d/\d t$ and the subscript $u$ for the radial
    coordinate derivative $\d/\d x^1$. One can notice that \eq(\ref{st2})
    decouples from perturbations other than $\delta\psi$ if one chooses the
    frame of reference and the coordinates in the perturbed space-time
    (the gauge for short) so that
\beq
    \delta\alpha = 2\delta\beta + \delta\gamma.           \label{st3}
\eeq
    The relation $\alpha=2\beta + \gamma$ thus holds for both the static
    background written as in (\ref{psi}), (\ref{ds-E}) and the
    perturbations. The unperturbed part of \eq (\ref{st2}) reads
    $\psi_{uu}=0$ and is satisfied by (\ref{psi}), while for $\delta\psi$ we
    obtain the wave equation
\beq
         \e^{4\beta(u)} (\delta\psi)\,\ddot{} - \psi_{uu}=0.      \label{st4}
\eeq
    The static nature of the background solution makes it possible to
    separate the variables,
\beq
    \delta\psi = \Phi(u) \e^{i\omega t},                      \label{Phi}
\eeq
    and to reduce the stability problem to a boundary-value problem for
    $\psi(u)$. Namely, if there exists a nontrivial solution to
    (\ref{st4}) with $\omega^2 <0$, satisfying some physically
    reasonable boundary conditions, then the static background system is
    unstable since the perturbations can exponentially grow with $t$.
    Otherwise it is stable in the linear approximation.

    Suppose $-\omega^2 = \Omega^2,\ \Omega > 0$. The equation that follows
    directly from (\ref{st4}),
\beq
    \Phi_{uu} - \Omega^2 \e^{4\beta(u)}\Phi=0,              \label{ePhi}
\eeq
    is converted to the normal Liouville (Schr\"odinger-like) form
\bearr
    d^2 y/dx^2 - [\Omega^2+V(x)] y(x) =0,   \nnn \cm
    V(x) = \e^{-4\beta}(\beta_{uu}-\beta_u{}^2).            \label{schrod}
\ear
    by the transformation
\beq
    \Phi(u) = y(x)\e^{-\beta},\qquad                        \label{u2x}
                            x = - \int \e^{2\beta(u)}du.
\eeq

    \eq (\ref{schrod}) makes it possible to use the experience of
    quantum mechanics (QM): $\Omega^2$ here corresponds to $ - E$ in the
    Schr\"{o}dinger equation. In other words, the presence of ``negative
    energy levels'' $E= - \Omega^2 <0$ for the potential $V(x)$ indicates
    the instability of our system.

    $V(x)$ is written explicitly as a function of $u$:
\beq                                                         \label{V}
    V(u)=-\frac{\sinh^3(2hu)}{32 h^2}
                          \left(\e^{-2hu} - 9\e^{-6hu}\right).
\eeq
    The variable $x$ behaves as follows at small and large $u$:
\begin{description}
\item [] $u\to 0$ (spatial infinity): $x\approx \e^\beta\approx 1/u$;
\item [] $u\to \infty$ (the sphere \Str): $x\approx 8h \e^{-2hu}$.
\end{description}
    For the potential $V(x)$ one finds:
\bearr
     V(x) \approx 2h/x^3 \quad
            (x\to\infty \quad \mbox{--- spatial asymptotic}),
\nnn
     V(x) \approx -1/(4x^2) \
                (x\to 0 \ \mbox{--- the sphere \Str.}).
\ear
    Thus we have a quadratic potential well at \Str, which is placed at $x=0$
    by choosing the proper value of the arbitrary constant in the definition
    of $x$ in \eq (\ref{u2x}).

    The same form of \eq (\ref{schrod}) with the same potential $V$ is
    obtained for the Einstein frame $\oM'$, the other part of the \wh, due
    to equal values of the integration constants in $\oM$ and $\oM'$.
    It makes sense, however, to change $x\to -x$, which does not affect \eq
    (\ref{schrod}) but makes it possible to unify the perturbation equations
    for the two parts of $\M$, the space-time of the Jordan-frame. We thus
    have \eq (\ref{schrod}) with $x \in \R$ and an even function
    $V(x)=V(-x)$, providing a potential well of the form $V\approx 1/(4x^2)$
    near $x=0$.

    The boundary conditions at both spatial asymptotics are obtained from
    the requirement that the perturbations should possess finite energy.
    This requirement upon the perturbed EMT leads to the condition
    $xy\to 0$ as $x\to\pm\infty$. Meanwhile, the
    asymptotic form of any solution of (\ref{schrod}) with $\Omega >0$
    at large $|x|$ is
\beq                                                          \label{asymp}
    y\approx C_1\e^{\Omega |x|}+C_2\e^{-\Omega |x|}, \qquad
    C_{1,2} = \const.
\eeq
    Therefore an admissible solution is the one with $C_1=0$, with
    only a decaying exponential. Actually, the conditions at both
    infinities are that $y\to 0$, i.e., coincide with the boundary
    conditions for the one-dimensional wave function under the
    same potential in QM.

    As is evident from QM (see, e.g., \cite{Rybakov}), a potential well of
    the form $V\approx 1/(4x^2)$ always possesses negative energy levels,
    $E = -\Omega^2 < 0$; moreover, the absolute value of $\Omega$ has no
    upper bound. The latter statement can be proved, e.g., by comparing
    \eq (\ref{schrod}) with our $V(x)$ and with rectangular potentials
    $\tilde V \geq V$ for which $y(x)$ and $E$ are easily found; one can then
    use the fact that $E_{\min}[V] < E_{\min}[\tilde V]$ where $E_{\min}$ is
    the lowest energy level (ground state) for a given potential.

    Recalling that $\Omega$ is the perturbation growth increment, we can
    conclude that our \wh{}s decay instantaneously within linear
    perturbation theory. Nonperturbative analysis would probably smooth out
    this infinite decay rate.

    It is of interest to note that at small $x$ the scalar field
    perturbation $\delta\phi$ behaves as $y\sqrt{|x|}$, hence the smoothness
    of the ``wave function'' $y(x)$ at $x=0$ implies $\delta\phi(0) =0$.
    In other words, the perturbation rapidly grows around \Str\ but vanishes
    on this sphere itself.

\section{Concluding remarks}

    Our perturbation analysis proves the instability of the \wh\ solutions
    for any $\xi > 0$. This violent instability occurs due to a negative
    pole of the perturbational effective potential $V(x)$ at the
    sphere \Str\ where vanishes the original function $f(\phi)$ in the action
    (\ref{act}), which actually means that the effective gravitational
    constant, proportional to $f^{-1}$, blows up and changes its sign.

    The instability of black holes with a conformal scalar field, found
    long ago in \Ref {br78}, is another example of such a phenomenon.
    It is quite plausible that instabilities of this kind are a common
    feature of STT solutions with conformal continuations, such that the
    transformation (\ref{trans-g}) maps the Einstein-frame manifold
    $\oM[\og]$ to only a part of the whole Jordan-frame manifold $\M[g]$
    and after which the effective gravitational coupling becomes negative.
    A similar instability was pointed out by Starobinsky \cite{star81} for
    cosmological models with conformally coupled scalar fields.
    We hope to return to this subject in our future work.

\Acknow{We are grateful to Vladimir Ivashchuk, Vitaly Melnikov, Yuri
    Rybakov, Georgy Shikin and Alexei Starobinsky for helpful discussions.
    We acknowledge partial financial support from the Ministry of Industry,
    Science and Technologies of Russia and Russian Basic Research
    Foundation.}

\end{document}